\documentclass{article}
\usepackage{spconf,amsmath,graphicx}
\usepackage{gensymb}
\usepackage{siunitx}
\usepackage{multirow}
\usepackage{enumitem}
\usepackage{footnote}
\usepackage[symbol]{footmisc}

\setlist{nosep, leftmargin=14pt}
\makesavenoteenv{tabular}

\usepackage{mwe} 
\usepackage[font=small,skip=0pt]{caption}

\title{4D iterative reconstruction of brain fMRI in the moving fetus}
 \name{\begin{tabular}{c}Athena Taymourtash$^{1}$, Hamza Kebiri$^{2,3}$, Sébastien Tourbier$^{2}$, Ernst Schwartz$^{1}$, Karl-Heinz Nenning$^{1}$, \\
 Roxane Licandro$^{1}$, Daniel Sobotka$^{1}$, Hélène Lajous$^{2,3}$,Priscille de Dumast$^{2,3}$, \\
 Meritxell Bach Cuadra$^{*,2,3}$, and Georg Langs$^{*,1}$\thanks{* these authors contributed equally to this work.}
 \end{tabular}
 }

 \address{$^{1}$ Computational Imaging Research Lab, Department of Biomedical Imaging and \\ Image-guided Therapy,Medical University of Vienna, Vienna, Austria. \\
 $^{2}$  Department of Radiology, Lausanne University Hospital (CHUV) and \\ University of Lausanne (UNIL), Lausanne, Switzerland. \\
 $^{3}$ CIBM Center for Biomedical Imaging, Switzerland.}

\begin{document}
%
\maketitle
\begin{abstract}
Resting-state functional Magnetic Resonance Imaging (fMRI) is a powerful imaging technique for studying functional development of the brain \textit{in utero}. However, unpredictable and excessive movement of fetuses has limited clinical application since it causes substantial signal fluctuations which can systematically alter observed patterns of functional connectivity. Previous studies have focused on the accurate estimation of the motion parameters in case of large fetal head movement and used a 3D single step interpolation approach at each timepoint to recover motion-free fMRI images. This does not guarantee that the reconstructed image corresponds to the minimum error representation of fMRI time series given the acquired data. Here, we propose a novel technique based on four dimensional iterative reconstruction of the scattered slices acquired during fetal fMRI. The accuracy of the proposed method was quantitatively evaluated on a group of real clinical fMRI fetuses. The results indicate improvements of reconstruction quality compared to the conventional 3D interpolation approach.
\end{abstract}
\section{Introduction}

Functional magnetic resonance imaging (fMRI) offers a unique means of observing the functional brain architecture and its variation during development, aging, or disease. Since its invention in the early 1990s, fMRI has revolutionized neuroscience by characterizing functional connectivity (FC) as the correlation between regional brain activity over time. Despite the insights into network formation and functional growth of the brain, \textit{in utero} fMRI of living human fetuses, however, remains challenging.
Unconstrained and potentially large movements of the fetuses during fMRI acquisition result in artifacts such as in-plane image blurring, slice cross-talk and spin-history artifacts that can considerably affect the image quality and bias any subsequent conclusions on the FC of the developing brain. 

Some of the early works in this area established tools from preprocessing of adult fMRI data (eg. MCFLIRT \cite{jenkinson2002improved}) and censored parts of the data with excessive motion (scrubbing). The reported rejection rate in these studies is very high and many subjects were removed since there was not enough data left after scrubbing. Further attempts to correct for fetal fMRI motion-related artifacts mainly relied on improving the estimate of the realignment parameters for potentially large movement of the fetuses. Ferrazzi et al. \cite{ferrazzi2014resting} proposed a comprehensive processing framework that incorporates bias field and spin-history correction, combined with slice to volume registration and scattered data interpolation, which allowed the data to be analyzed within a single consistent space . Furthermore, to avoid that surrounding maternal tissues dominate the similarity metric of registration algorithms, \cite{rutherford2019observing} trained a convolutional neural network for rapid and accurate isolation of the fetal brain and then utilized rigid registration for motion correction. You et al. also in \cite{you2016robust} present an optimized design based on the MCFLIRT registration to identify independently moving objects (i.e., the placenta and fetal brain) and estimate the local motion for each. All these studies, however, used 3D volume interpolation of each time frame independently to reconstruct the whole FC data. More recently, \cite{sobotka2019reproducibility} extended super resolution technique (will be discussed in the next section) with separate reconstruction for each 3D volume within the time sequence. Since in-utero motion is unconstrained and complex, the rate of scattered points versus the points to infer is less than 1 to 1 in each 3D volume. Therefore, taking the advantage of temporal structure in fMRI imaging can potentially improve the reconstruction of image in this challenging problem. Here we propose a new method that, rather than treating each volume independently, takes both spatial- and the temporal domain into account and iteratively reconstruct 4D fetal fMRI time series.

\section{Method}

Our method directly reconstructs fetal fMRI time series on a 4D regular grid from motion scattered slices. We formulate our problem as a maximum likelihood (ML) estimation in which the conditional probability density function (pdf) of the acquired slices given the current estimate of the reconstruction is maximized: 
\begin{equation}
\begin{split}
\widehat{\mathbf{X}} & = \arg \max _{\mathbf{X}}\prod_k \text{Pr}(\mathbf{X}|S_k) \\
& = \arg \max _{\mathbf{X}}\prod_k \text{Pr}(S_k|\mathbf{X}) \frac{\text{Pr}(\mathbf{X})}{\text{Pr}(S_k)}.
\label{eq1}
\end{split}
\end{equation}
Here, $\widehat{\mathbf{X}}$ denotes an estimate of the final 4D reconstructed image i.e. $\mathbf{X}$, and $S_k$ is the k-th acquired slice. $\text{Pr}(\mathbf{X})$ is a prior on $\widehat{\mathbf{X}}$ and $\text{Pr}(S_k)$ is a constant with respect to $\widehat{\mathbf{X}}$. Using the assumption of Gaussian noise with zero mean and standard deviation of $\sigma_k$, and independent acquisition of each slice, the above conditional pdf can be written as
\begin{equation}
\text{Pr}(S_k|\widehat{\mathbf{X}})=\prod_i\frac{1}{\sigma_k\sqrt{2\pi}}\exp(-\frac{(\widehat{S}_k(i)-S_k(i))^2}{2\sigma_k^{2}}),
\label{eq2}
\end{equation}
where $S_k(i)$ are the samples from the acquired slices $S_k$, and $S_k(i)$ are the samples from the estimated slices by considering a slice acquisition model as $
\mathbf{H}_{k}=\mathbf{D}_{k} \mathbf{B}_{k} \mathbf{M}_{k}+\mathbf{n}_{k}
$. $\mathbf{D}_{k}$ is the subsampling matrix, $\mathbf{B}_{k}$ is the blur matrix representing the point spread function (psf), and $\mathbf{n}_{k}$ is the observation noise. The motion matrix $\mathbf{M}_{k}$ is defined for each slice through slice-to-volume registration as a 6-DOF 3-D rigid transformation (including three rotations and three translations). When the noise residuals are presumed to be drawn from a Gaussian distribution, and using the logarithmic function, the above problem is equivalent to obtaining the Maximum A posteriori Probability (MAP) estimate by minimizing the log likelihood function:
\begin{equation}
\widehat{\mathbf{X}} = \arg \min _{\mathbf{X}} \lambda \sum_{k}\left\|\mathbf{H}_{k} \mathbf{X}-S_{k}\right\|^{2}+ \frac{\alpha}{2}R(\mathbf{X})
\end{equation}
The ML formulation is applied in structural fetal T2-weighted MRI where a high-resolution volume is reconstructed using several orthogonal series of thick slices, called as super-resolution technique \cite{gholipour2010robust,kuklisova2012reconstruction,tourbier2015efficient,ebner2020automated}. The point spread function used in that case is derived directly from the image acquisition process. In fMRI analysis, however, low-frequency fluctuations of the signal over time is desired and post-processing commonly includes low-pass filtering to derive temporally smooth signal. In this work we propose, rather than using a separate post-processing filter, to integrate this step within the reconstruction framework by using a four dimensional point spread function. This also provides further stability of the reconstruction by incorporating information from the neighboring slices in time in case of a large motion and missing data. The 4D psf can be formulated as:
\begin{equation}
G(x,y,z,t) = \frac{1}{2\pi\sigma^{2}}\exp{^{(\frac{x^2}{2\sigma_x^{2}}+\frac{y^2}{2\sigma_y^{2}}+\frac{z^2}{2\sigma_z^{2}}+\frac{t^2}{2\sigma_t^{2}})}}
\label{eq4}
\end{equation}
where the $\sigma$ values represent the point spread in x, y, z, and t,  which are the neighborhood of a voxel is defined on both spatial and temporal domain. To create a distance metric in space–time, we scaled the time dimension, as suggested in \cite{ferrazzi2014resting}, by a factor
of \(\frac{\textrm{res(z)}}{\textrm{TR}}\), where res(z) is the through-plane resolution of the image and TR is the repetition time.

The algorithm involves iterative backprojection of the motion scattered slices onto a regular 4D grid using a 4D psf, estimating the slice intensities given the current estimate of the 4D image (forward projection using 4D psf), and computing the residual error between these two. This error is backprojected and added to the regularization term which was computed using a first order Tikhonov function. Optimization terminates when the residual error between estimated and measured data converges. 

\begin{figure}[t]
\begin{minipage}[b]{1.0\linewidth}
  \centering
  \centerline{\includegraphics[width=8.5cm]{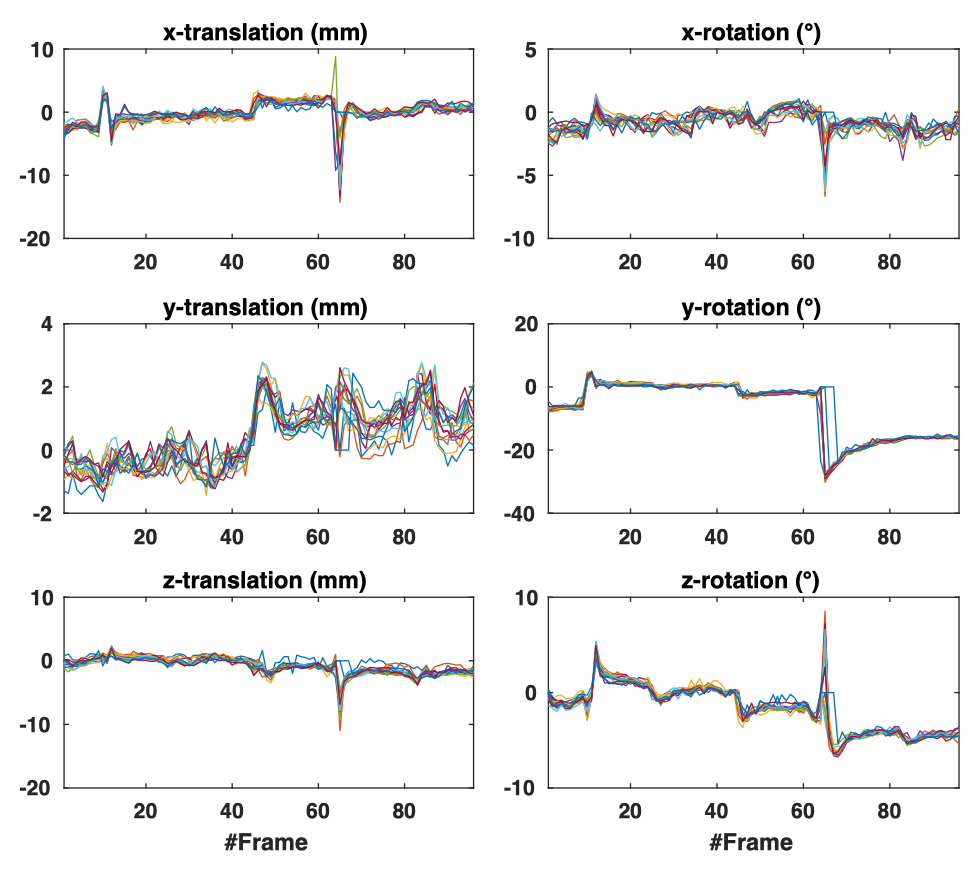}}
\end{minipage}
\caption{Example of slice-wise head motion parameters estimated for one fetus with strong rotational motion using a symmetric block matching registration algorithm based on the normalized cross correlation including three rotations in degrees and three translations in millimeters.}
\label{fig:realign}
\end{figure}

\subsection{Implementation details}
We implemented our algorithm of 4D reconstruction based on the publicly available toolkit of Numerical Solver Library (NSol) which is a part of NiftyMIC package \cite{ebner2020automated} for volumetric reconstruction of the structural 3D images. A binary brain mask was manually delineated on the average volume of each fetus and dilated to ensure it covers the fetal brain through all ranges of the motion. A four dimensional estimate of the bias field for spatiotemporal signal non-uniformity correction in fMRI series was obtained using N4ITK algorithm \cite{tustison2010n4itk} as suggested previously \cite{seshamani2014method}. The motion correction step involved globally co-registering all time points, creating a target volume as proposed in \cite{seshamani2013cascaded} by automatically finding a set of consecutive volumes of fetal quiescence and averaging over them, and finally performing hierarchical slice-to-volume registration based on the interleaved factor of acquisition as implemented in NiftyMIC package\cite{ebner2020automated}.

\begin{figure}[t]
\begin{minipage}[b]{1.0\linewidth}
  \centering
  \centerline{\includegraphics[width=8.5cm]{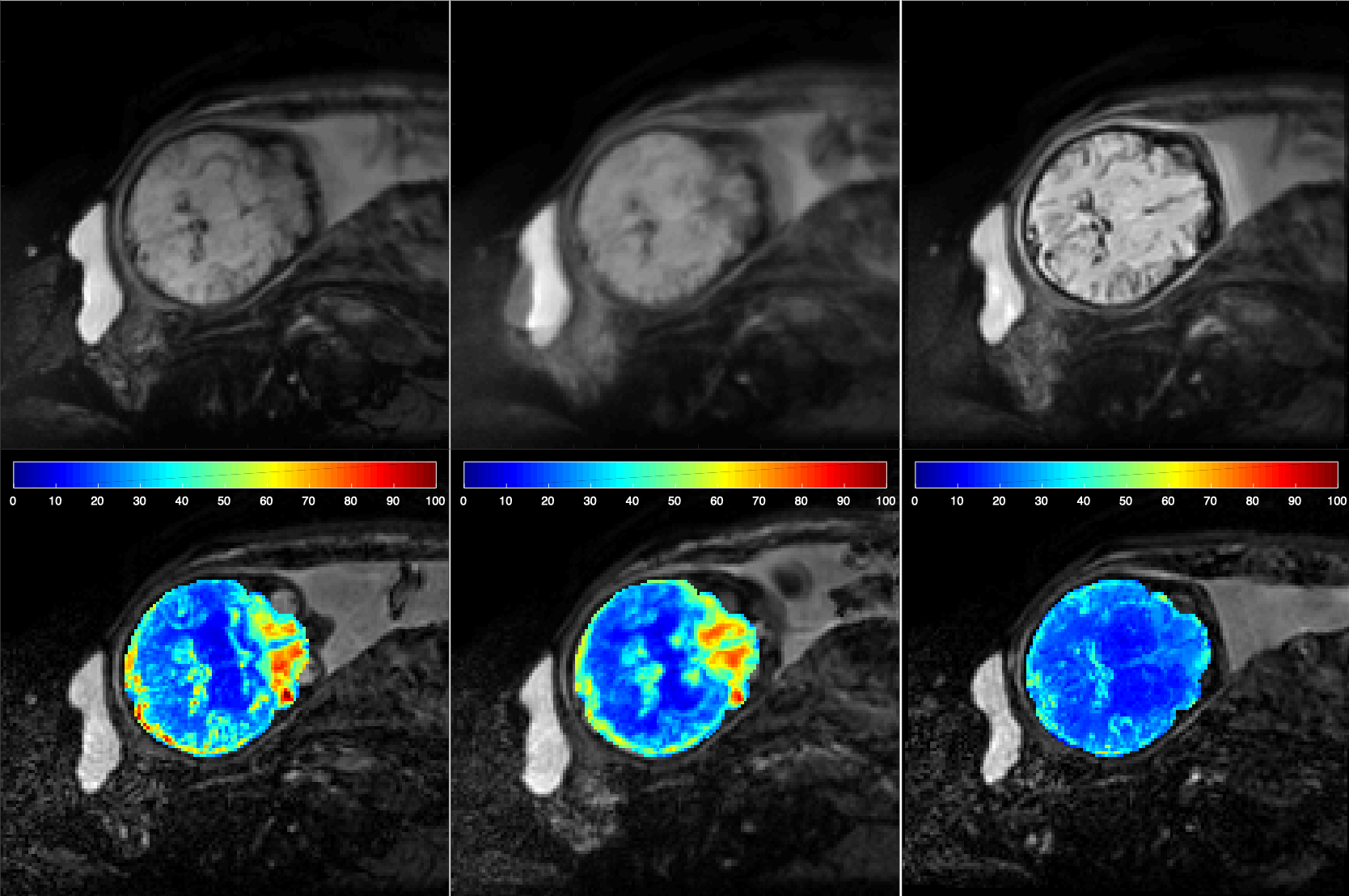}}
  \centerline{\includegraphics[width=8.5cm]{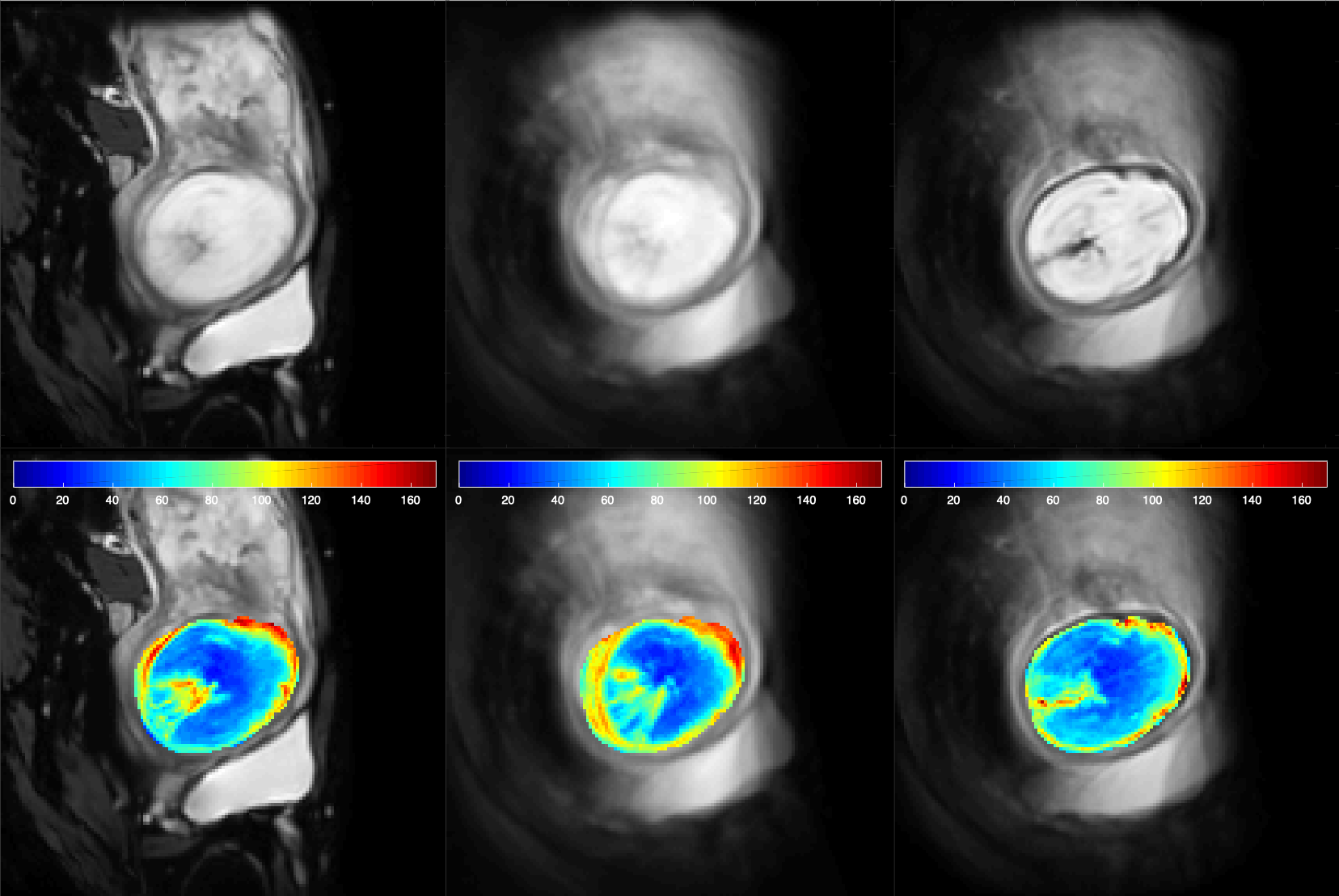}}
\end{minipage}
\caption{Examples in the average volume (top row) and voxelwise standard deviation of the signal (bottom row) for two fetal subjects of \textbf{S14} and \textbf{S8} whit strong rotational parameters. The columns correspond to the original data, single step 3D linear interpolation, and the proposed 4D iterative reconstruction based on the maximum a-posteriori estimation minimizing a first order Tikhonov function.}
\label{fig:recon}
\end{figure}

\section{Experimental Results}
We evaluated the proposed algorithm on fetal fMRI scans of 15 subjects between 22  and 39 weeks of gestation (Table.\ref{tab:vae}). None of the cases showed any neurological pathologies. Pregnant women were scanned on a 1.5T clinical scanner (Philips Medical Systems, Best, Netherlands) using single-shot echo-planar imaging (EPI), and a sensitivity encoding (SENSE) cardiac coil with five elements. Image matrix size was 144$\times$144, with 1.74$\times$1.74 $mm^{2}$ in-plane resolution, 3$mm$ slice thickness, flip angle of 90\si{\degree}, and 96 volumes per acquisition. We used a hierarchical approach based on a symmetric block matching registration algorithm to estimate the rigid realignment parameters of each slice. Table.\ref{tab:vae} summarizes the estimated motional parameters for all subjects, showing especially free excessive rotation in fMRI images of the fetal population. Figure 1 shows the slice-wise realignment parameters for a subject with strong rotation (27.9\degree in left-right direction). When such big motion occurs, gaps might be open up in the reconstructed image between slices that are spatially adjacent. Therefore, interpolation methods using only spatial information, although faster in implementation, may not correct it as there would be no data contributing to the reconstruction of a given regular grid. As fMRI data is temporally correlated, adjacent slices in time can potentially add information to the reconstruction. Here we used a four dimensional Gaussian kernel as psf to take both spatial and temporal structure of the data into account, and iteratively optimized our reconstruction through inverse problem framework.  

\begin{figure}[!b]
\begin{minipage}[b]{1.0\linewidth}
  \centering
  \centerline{\includegraphics[width=8.5cm]{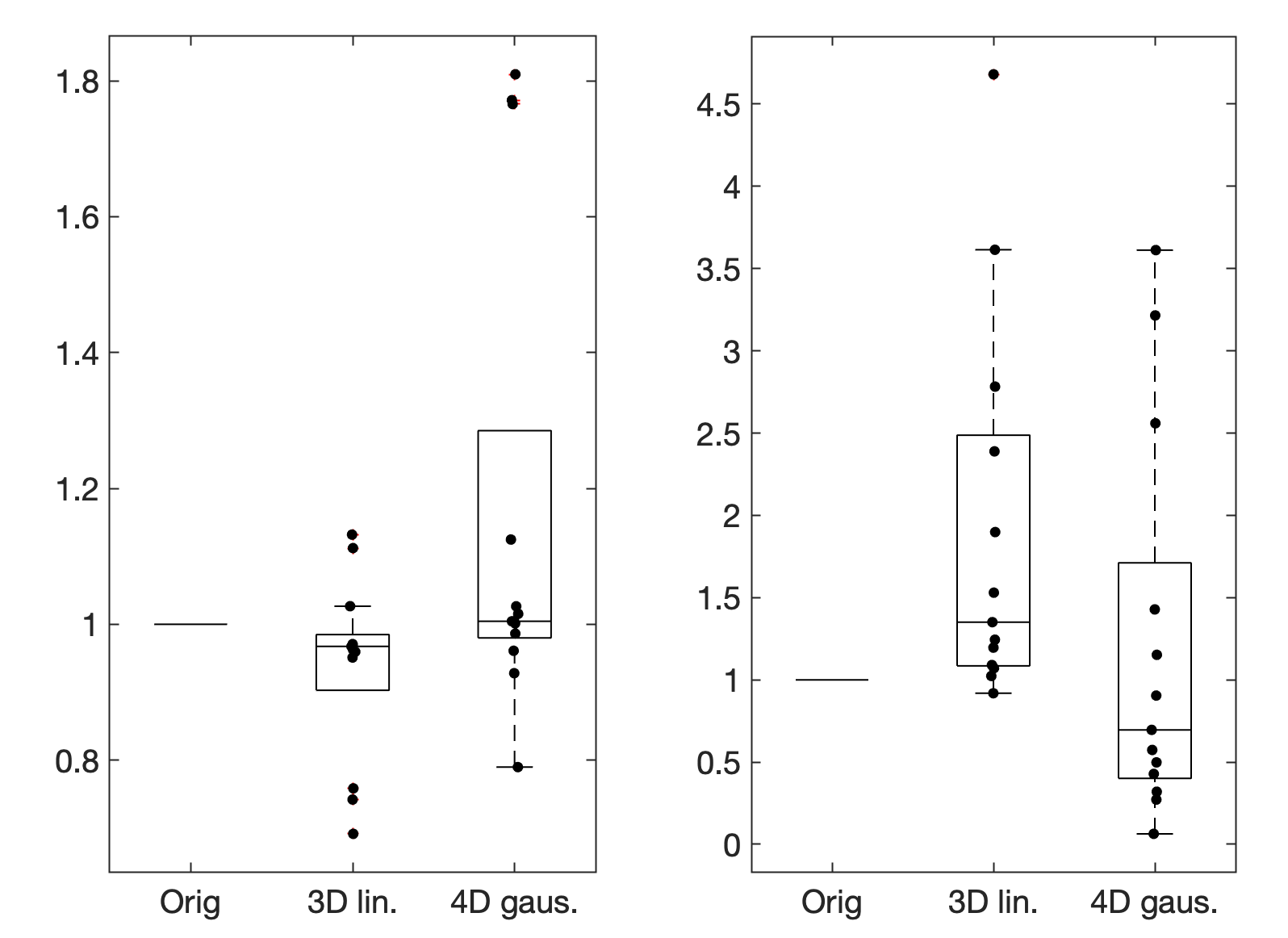}}
\end{minipage}
\caption{boxplot analysis and statistics of the sharpness (left) and standard deviation of BOLD signal fluctuation (right) for 15 fetuses.}
\label{fig:box}
\end{figure}

\begin{table*}[t]
\centering
\small
    {\caption{Motion characteristics, sharpeness and signal intensity standard deviation for all subjects.}\label{tab:vae}} %
    \resizebox{\linewidth}{!}{
    {\begin{tabular}{l c l l l l l l || c c c |c c c}
    \hline
             &  \textbf{GA} & \multicolumn{3}{c}{\textbf{Max translation (mm)}} & \multicolumn{3}{c|| }{\textbf{Max rotation (\textbf{\degree)}}} & \multicolumn{3}{c|}{\textbf{Sharpness}} &\multicolumn{3}{c}{\textbf{Standard deviation}}\\
             &  \textbf{(week+day)} & \textbf{x}&\textbf{y}&\textbf{z} & \textbf{roll}&\textbf{pitch}& \textbf{yaw} & \textbf{Raw} &\textbf{Linear}& \textbf{Ours} &\textbf{Raw}& \textbf{Linear}& \textbf{Ours}\\
    \hline
    \hline
      \textbf{S1} & 22w6d & 4.1 & 6.1& 11.4 & 25.2& 28.2& 9.5 & \textbf{16949} & 11721&15729& \textbf{1.60} &7.49 & 5.78\\ \hline
     \textbf{S2} & 23w4d &  5.3& 7.4& 8 & 6.8& 3.3& 3.4 & \textbf{20409} & 15143 & 16119 & \textbf{18.87} & 45.08 & 48.30 \\ \hline
    \textbf{S3}  & 25w0d & 1.6& 9.1& 3.8 & 8.2& 0.7& 1.9 & 4672  & 5196 &\textbf{5255} & \textbf{6.37} & 6.82 & 7.34 \\ \hline
    \textbf{S4}  & 25w4d & 2.9& 3.4& 5 & 5.7& 7.6& 9.1 & 30539 & \textbf{31353} & 31010 & 18.59 & 35.29 & \textbf{5.06} \\ \hline
    \textbf{S6}  & 28w6d & 2.5& 1.7& 7.6 & 3.4& 6.3& 1.5 & 30855 & 28746 & \textbf{31298} & 16.34 & 25.03 & \textbf{12.22} \\ \hline
    \textbf{S7}  & 29w3d  & 20.7& 6& 6.2 & 5.4& 17.1& 9.5 & 32761 & 31158 & \textbf{32807} & \textbf{19.37} & 53.90 & 62.28 \\ \hline
    \textbf{S8}  & 29w4d & 2.7& 3.8& 2.6 & 16.4& 14.7& 21.6 & 32299 & 30993 & \textbf{33154} & 43.07 & 58.17 & \textbf{21.75}\\ \hline
    \textbf{S9}  & 29w5d  & 4.2& 5.8& 2.7 & 5.6& 4.5& 3.4 & 33218 & 32136 & \textbf{33368} & 46.41 & 71.01 & \textbf{19.87} \\ \hline
    \textbf{S10}  & 30w2d  & 4.7& 5.6& 4.3 & 14.3& 2& 21.6 & \textbf{22531} & 21854 & 21653 & 68.15  & 74.28 & \textbf{21.79}\\ \hline
    \textbf{S11}  & 32w2d  & 9.2& 5.2& 4.8 & 8.4& 18.9& 12.5 & 33218 & 32136 & \textbf{33368} & 8.16 & 8.35 & \textbf{4.68}\\ \hline
    \textbf{S12}  & 34w4d  & 2.3& 5& 8.3 & 17.6& 12.2& 9.8 & \textbf{33198} & 32230 & 32744 & \textbf{22.42} & 81.04 & 32.02 \\ \hline
    \textbf{S13}  & 34w6d  & 2.8& 1.5& 3.8 & 1.4& 4.6& 1.7 & 696 & 670 & \textbf{1233} & 9.86 & 11.79 & \textbf{6.86} \\ \hline
    \textbf{S14}  & 35w6d  & 11.9& 3& 3.2 & 4.2& 27.9& 7.7 & 1035 & 785 & \textbf{1873} & 16.44 & 15.10 & \textbf{8.19}\\ \hline
    \textbf{S15}  & 39w2d  & 1.4& 3.7& 1.5 & 3.6& 3.8& 1.3 & 803 & 909 & \textbf{1418} & 8.89 & 11.06 &\textbf{8.04} \\ \hline
    \hline
    \end{tabular}}}
\end{table*}

\subsection{Reconstruction accuracy}
We compared the proposed 4D reconstruction with standard 3D linear reconstruction. We evaluated the sharpness  \cite{pech2000diatom} of the averaged volume after motion correction and reconstruction, 
and the standard deviation of BOLD signal fluctuations at each voxel over time. Figure \ref{fig:recon} illustrates the average volume (top row), and standard deviation of intensity over time (bottom row) of the original data without motion correction, data after single step 3D linear interpolation, and data reconstructed using the proposed algorithm for two subjects with different motion characteristics. Severe blurring is observed in the image obtained from 3D linear interpolation. In contrast, the proposed method yields qualitatively sharper images as the effect of slice motion is appropriately decreased. Quantitative results are reported in Table~\ref{tab:vae} for all 15 subjects. Figure \ref{fig:box} shows the boxplot for the relative changes of the sharpness (left) and standard deviation (right) in 3D interpolation vs 4D Gaussian reconsruction normalized to the corresponding value in raw data.  In 12 out of 14 subjects the proposed method yields higher sharpness than 3D interpolation, with an overall average sharpness $26\%$ higher than for 3D interpolation, in 11 it exhibits lower standard deviation, with an average standard deviation reduction of $41\%$ compared to 3D interpolation. For several subjects with lower gestational age the highest sharpness and lowest standard deviation occurs with un-corrected images. One possible cause is that the advanced gyrification at later gestational ages results in a pronounced sharpness advantage with correction, while for younger fetuses this is not the case and might reflect image noise rather than anatomical structure. 

\section{Conclusion}
In this work we proposed a novel spatio-temporal reconstruction approach for the fetal brain acquired while there is unconstrained motion of the head.  Our technique integrates the temporal continuity of fMRI time series with the spatial coherency of the neighboring brain voxels by using a four dimensional Gaussian point spread function for reconstructing motion scattered slices. Qualitative results of the resulting reconstruction of clinical fetal fMRI data shows improvement by the method. Quantitative evaluation confirms this improvement predominantly for higher gestational age. As the developed formulation for fMRI time series reconstruction is general and can be effectively used in adult studies as well, it would be of interest to investigate its performance on the adult population in the future.

\section{Acknowledgment}
This work has received funding from the European Union's Horizon 2020 research and innovation programme under the Marie Sk\l odowska-Curie grant agreement No 765148, and partial funding from the Austrian Science Fund (FWF, P35189).
This work was also supported by the Swiss National Science Foundation (project 205321-182602). We acknowledge access to the expertise of the CIBM Center for Biomedical Imaging, a Swiss research center of excellence founded and supported by Lausanne University Hospital (CHUV), University of Lausanne (UNIL), Ecole polytechnique fédérale de Lausanne (EPFL), University of Geneva (UNIGE) and Geneva University Hospitals (HUG).

\bibliographystyle{IEEEbib}
\bibliography{main}

\end{document}